\documentclass[12pt,draftclsnofoot,onecolumn]{IEEEtran}

\usepackage{amsmath,amssymb,amsfonts}
\usepackage{amsthm}
\theoremstyle{definition}

\usepackage{soul}
\newtheorem{proposition}{Proposition}
\newif\ifappendices
\appendicestrue
\newif\iflongversion
\longversionfalse
\usepackage{algorithmic}
\usepackage{graphicx}
\usepackage{textcomp}
\usepackage{xcolor}
\newcommand{\rev}[1]{#1}
\usepackage{bm}
\usepackage{url}
\usepackage{pgfplots}
\pgfplotsset{compat=1.17}
\usepgfplotslibrary{groupplots}

\usepackage[backend=biber, style=ieee, sorting=none]{biblatex}
\graphicspath{{./}}

\pgfplotsset{
  fasbase/.style={
    width=0.52\textwidth, height=0.40\textwidth, ymode=log,
    grid=both, grid style={gray!18}, unbounded coords=discard,
    tick label style={font=\footnotesize}, label style={font=\footnotesize},
    title style={font=\footnotesize}, legend cell align=left,
    legend style={font=\scriptsize, fill opacity=0.85, draw=none},
  },
  every axis plot/.append style={line width=0.8pt},
}
\newcommand{\Kone}{blue}
\newcommand{\Ktwo}{orange}
\newcommand{\Kfour}{green!60!black}
\newcommand{\Keight}{red}

\newcommand{\mat}[1]{\bm{#1}}           
\renewcommand{\vec}[1]{\bm{#1}}         
\DeclareMathOperator{\E}{\mathbb{E}}    

\DeclareMathOperator*{\argmax}{arg\,max}

\usepackage[acronym,nonumberlist,nopostdot,nogroupskip]{glossaries}
\newacronym[shortplural={FAS},longplural={fluid antenna systems}]%
    {fas}{FAS}{fluid antenna system}
\newacronym{oodn}{OODN}{on-off digital noise}
\newacronym{iot}{IoT}{Internet of Things}
\newacronym{awgn}{AWGN}{additive white Gaussian noise}
\newacronym{bep}{BEP}{bit error probability}
\newacronym{snr}{SNR}{signal-to-noise ratio}
\newacronym{lrt}{LRT}{likelihood-ratio test}
\newacronym{cdf}{CDF}{cumulative distribution function}
\newacronym{pdf}{PDF}{probability density function}
\newacronym{fama}{FAMA}{fluid antenna multiple access}
\newacronym{los}{LoS}{line-of-sight}
\newacronym{nlos}{NLoS}{non-line-of-sight}
\newacronym{mc}{MC}{Monte Carlo}

\begin{document}

\title{On-Off Digital Noise Modulation with Fluid Antenna Systems over $\kappa$-$\mu$ Fading Channels}

\author{%
    Daniel C.~Ara\'ujo,
    Andr\'e A.~dos~Anjos,
    Hugerles S.~Silva,
    Constantinos Psomas,
    and Robson D.~Vieira%
}

\maketitle

\begin{abstract}
This letter proposes the integration of on-off digital noise~(OODN) modulation with fluid antenna systems~(FASs). A unified analytical framework is developed to evaluate the performance of FAS-assisted OODN receivers over additive white Gaussian noise and generalized $\kappa$-$\mu$ fading channels. The analysis incorporates fluid antenna port selection, the number of available ports, and spatial correlation. Analytical expressions for the average bit error probability are derived and the achievable diversity order is characterized. All expressions are validated through Monte Carlo simulations. Results demonstrate that the spatial diversity provided by the FAS significantly enhances the reliability of OODN transmissions while preserving their inherent low-complexity, non-coherent operation without carrier-phase recovery. The proposed framework establishes a new research direction for energy-efficient Internet of Things (IoT) and machine-type communication systems.
\end{abstract}


\IEEEpeerreviewmaketitle

\section{Introduction}

\IEEEPARstart{N}{ext}-generation wireless networks are expected to simultaneously support massive numbers of low-cost and energy-constrained devices while providing reliable connectivity under increasingly crowded spectrum conditions~\cite{Chen}. These requirements have stimulated the development of communication techniques that reduce hardware complexity, minimize synchronization requirements, and improve resilience against fading and interference. In this context, two emerging research directions have recently attracted attention, namely noise-based modulation and~\glspl{fas}. Although both technologies pursue low-complexity and energy-efficient wireless communications, they have evolved independently and their joint potential remains unexplored.

Noise-based communication has progressed considerably from the early concepts of Johnson-noise and thermal-noise communications toward practical digital modulation schemes. Recent advances include modulated Johnson-noise transmitters, thermal-noise communication, digital noise modulation, non-orthogonal multiple access, multidimensional noise signaling, and differential detection techniques~\cite{Kish_2005,Kish_2006,Kapetanovic_2022,Garman_2023,Basar2023TCOM,Basar2024,Yapici_2024,NoiseModSpread,NoiseMod3d,dbn_2026}. Among these techniques, \gls{oodn} modulation has recently emerged as a particularly attractive solution for \gls{iot} and machine-type communications by encoding information in the variance of a random waveform rather than on a deterministic carrier~\cite{dosAnjos2025}. \Gls{oodn} enables phase-noncoherent reception, avoiding carrier-phase and frequency synchronization while maintaining very low implementation complexity. Existing studies have characterized its detection performance over \gls{awgn} and fading channels~\cite{dosAnjos2025,dosanjos2}. Nevertheless, all available analyses assume single-antenna receivers.

In parallel, \glspl{fas} have recently emerged as a promising physical-layer architecture capable of exploiting spatial diversity through a single reconfigurable antenna that dynamically switches among multiple closely spaced ports~\cite{Wong2023}. By selecting the port with the most favorable propagation condition, a \gls{fas} provides diversity gains without requiring multiple radio-frequency chains, making it especially attractive for compact and low-power terminals. The literature has rapidly expanded from fundamental performance analyses over simple or generalized fading channels toward multi-user \gls{fama}, machine-learning-assisted port selection, and integrated communication scenarios~\cite{wong2020limits,FAMANew,LNNFAMA,isac}. Despite these advances, all existing \gls{fas} works assume coherent modulation, whereas its interaction with non-coherent noise-based signaling has never been investigated, to the best of our knowledge.

In this letter, we introduce and analyze for the first time the integration of a \gls{fas} and \gls{oodn}. We develop an analytical framework to characterize the performance of \gls{oodn} receivers employing a \gls{fas} under both \gls{awgn} and $\kappa$-$\mu$ fading channels. The proposed analysis quantifies the diversity gains achieved through fluid antenna port selection, investigates the influence of the number of available ports and spatial correlation, and reveals how the \gls{fas} fundamentally modifies the detection performance of noise-based modulation. Numerical results demonstrate that the spatial diversity offered by the \gls{fas} substantially improves the reliability of \gls{oodn} transmissions without compromising their inherent low-complexity and non-coherent characteristics.

\section{System Model}
\begin{figure}[!t]
    \centering
    \includegraphics[width=\columnwidth]{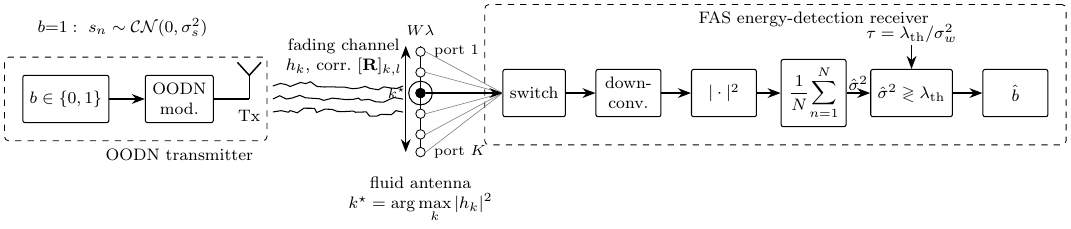}
    \caption{\Gls{oodn} transmitter and fluid-antenna energy-detection receiver.
    Bit~$1$ radiates a Gaussian noise burst and bit~$0$ is silent.}
    \label{fig:system}
\end{figure}

We consider a point-to-point link over a block-fading channel, where a
single-antenna transmitter sends a binary stream to a receiver equipped with a
fluid antenna, as shown in Fig.~\ref{fig:system}. Each information bit
occupies $N$ complex baseband samples of one coherence block, and the receiver
detects it non-coherently through an energy test.

The transmitter signals with \gls{oodn}~\cite{dosAnjos2025}, a member
of the noise-modulation family~\cite{Basar2024,dasilva2026survey}. It
maps the equiprobable bit $b\in\{0,1\}$ onto a burst of $N$ independent and identically distributed (i.i.d.) circularly
symmetric complex Gaussian samples $s_n=\tilde{s}_n$ for $b=1$ and $s_n=0$ for
$b=0$, with $\tilde{s}_n\sim\mathcal{CN}(0,\sigma_s^2)$ and $n=1,\dots,N$. Bit~$1$
therefore radiates a band-limited noise-like burst of
average power $\sigma_s^2$, while bit~$0$ leaves the channel silent, giving the
scheme its on-off nature. The waveform carries no deterministic structure, so
a coherent receiver gains nothing over an energy detector~\cite{urkowitz1967energy}.
This motivates the
square-law front end adopted below and underlies the low-probability-of-intercept
property of \gls{oodn}.

At the receiver, the fluid antenna repositions its single radiating element to
any one of $K$ preset ports uniformly distributed along a linear aperture of
length $W\lambda$, with $\lambda$ being the carrier wavelength. Let $h_k$ be the
flat-fading channel coefficient at port $k\in\{1,\dots,K\}$ and let
$\Omega_k=|h_k|^2$ be its normalized power of unit mean $\E[\Omega_k]=1$. We
model the small-scale fading as $\kappa$-$\mu$~\cite{yacoub2007kappamu}, a general
\gls{los} family whose unit-mean power density is
\begin{equation}
    f_{\Omega}(w) =
    \frac{\mu(1+\kappa)^{\frac{\mu+1}{2}}}{\kappa^{\frac{\mu-1}{2}}e^{\mu\kappa}}
    \, \frac{e^{-\mu(1+\kappa)w}}{w^{\frac{1-\mu}{2}}}\,
    I_{\mu-1}\!\bigl(2\mu\sqrt{\kappa(1+\kappa)w}\bigr),
    \label{eq:kappamu_pdf}
\end{equation}
for $w\ge0$, where $\kappa>0$ is the ratio of dominant to scattered power,
$\mu>0$ is the number of multipath clusters, and $I_{\nu}(\cdot)$ is the modified
Bessel function of the first kind of order $\nu$. The corresponding \gls{cdf} is
\begin{equation}
    F_{\Omega}(w) = 1 - Q_{\mu}\!\bigl(\sqrt{2\kappa\mu},\,
        \sqrt{2\mu(1+\kappa)\,w}\bigr),
    \label{eq:kappamu_cdf}
\end{equation}
with $Q_{\mu}(\cdot,\cdot)$ denoting the generalized Marcum-$Q$ function. This family
recovers Rayleigh fading at $\mu=1,\kappa\to0$, Rician at $\mu=1$, and
Nakagami-$m$ at $\kappa\to0,\mu=m$.

Because the ports share a small aperture, their gains are spatially correlated.
Under isotropic scattering, the reference correlation matrix
$\mat{R}\in\mathbb{R}^{K\times K}$ has Jakes' entries~\cite{jakes1974}
\begin{equation}
    [\mat{R}]_{k,l} = J_0\!\left(\frac{2\pi\,(k-l)}{K-1}\,W\right),
    \label{eq:jakes}
\end{equation}
where $J_0(\cdot)$ is the zeroth-order Bessel function of the first kind. Rather
than replace the full $\mat{R}$ by a partition into independent
blocks~\cite{ramirez2024blockcorr,demoura2026fama}, which is intractable to keep
exact~\cite{khammassi2023approx} and optimistic in the transition region, we
retain $\mat{R}$ exactly through a Gaussian copula. The $K$ spatially correlated
$\kappa$-$\mu$ port gains $\Omega_k=|h_k|^2$ are generated from a latent Gaussian vector
$\vec{z}=[z_1,\dots,z_K]^{\!\top}\sim\mathcal{N}(\vec 0,\mat{R})$ by matching
marginals,
\begin{equation}
    \Omega_k = F_{\Omega}^{-1}\!\bigl(\Phi(z_k)\bigr), \qquad k=1,\dots,K,
    \label{eq:copula}
\end{equation}
where $\Phi(\cdot)$ is the standard normal \gls{cdf} and $F_{\Omega}(\cdot)$ is the
marginal distribution of the $\kappa$-$\mu$ model, given by~\eqref{eq:kappamu_cdf}. The mapping~\eqref{eq:copula}
preserves the $\kappa$-$\mu$ marginal of every port exactly while imposing the
Jakes spatial dependence of~\eqref{eq:jakes}, so it reproduces the correlated
fluid-antenna channel without any block approximation. A smaller aperture $W$
raises the off-diagonal entries of $\mat{R}$ and couples the ports, reducing the
effective number of independent gains available to the port-selection rule that
picks the strongest port $k^\star$. A larger $W$ decorrelates the ports and,
through the oscillating lobes of $J_0$, can even render neighboring ports mildly
anti-correlated.

With the antenna at port $k$, the observed samples are
$x_{k,n}=h_k\,s_n+w_{k,n}$, where $w_{k,n}\sim\mathcal{CN}(0,\sigma_w^2)$ is the
\gls{awgn}, independent across ports and samples. We define the active-state
transmit \gls{snr} $\bar{\gamma}_S\triangleq\sigma_s^2/\sigma_w^2$ and the effective per-port
\gls{snr} $\gamma_{S,k}\triangleq|h_k|^2\sigma_s^2/\sigma_w^2=\Omega_k\,\bar{\gamma}_S$.
Bit $0$ is silent and the bits are equiprobable, so the average transmit \gls{snr}
is $\bar{\gamma}_S/2$ and all results are reported against $\bar{\gamma}_S$.

Following the fluid-antenna paradigm, the receiver activates the single port
that maximizes the channel gain, $k^\star=\argmax_{k\in\{1,\dots,K\}}|h_k|^2$. The
selected \gls{fas} port thus carries the largest normalized power
$\Omega_{\max}=\max_k\Omega_k=\Omega_{k^\star}$, and the effective \gls{snr} after
selection is $\gamma_S\triangleq\gamma_{S,k^\star}=\bar{\gamma}_S\,\Omega_{\max}$. By the spatial correlation in~\eqref{eq:jakes},
$\Omega_{\max}$ is the maximum of $K$ correlated $\kappa$-$\mu$ variates, and its
distribution sets the diversity gain over a fixed single-antenna \gls{oodn} link.

On the selected port, the detector forms the normalized energy statistic
$\hat{\sigma}^2\triangleq\frac{1}{N}\sum_{n=1}^{N}\bigl|x_{k^\star,n}\bigr|^2$
and applies the test
\begin{equation}
    \hat{\sigma}^2
    \underset{\hat{b}=0}{\overset{\hat{b}=1}{\gtrless}}
    \lambda_{\mathrm{th}} ,
    \qquad
    \tau \triangleq \lambda_{\mathrm{th}}/\sigma_w^2 ,
    \label{eq:decision}
\end{equation}
where \rev{$\lambda_{\mathrm{th}}$ is the energy threshold the statistic is
compared against,} $\hat b$ denotes the detected bit and $\tau$ the
noise-normalized threshold, a design parameter discussed in
Section~\ref{sec:analysis}.

Conditioned on the transmitted bit and the selected gain, the samples
$x_{k^\star,n}$ are i.i.d.\ $\mathcal{CN}\bigl(0,\sigma_w^2(1+\gamma_S
b)\bigr)$, so the scaled energy statistic that drives the
test~\eqref{eq:decision} is central chi-square with $2N$ degrees of freedom,
\begin{equation}
    \frac{2N\,\hat{\sigma}^2}{\sigma_w^2\,(1+\gamma_S b)}\sim\chi^2_{2N},
    \qquad b\in\{0,1\},
    \label{eq:stat_law}
\end{equation}
where $\gamma_S=\bar{\gamma}_S\,\Omega_{\max}$ is the selected effective \gls{snr}.
The selection variable $\Omega_{\max}=\max_k\Omega_k$ is the largest of the $K$
correlated gains, whose joint law follows from the copula~\eqref{eq:copula} with marginals $\kappa$-$\mu$ distributions given by~\eqref{eq:kappamu_cdf} and reference matrix $\mat{R}$
of~\eqref{eq:jakes}. Because the marginal map in~\eqref{eq:copula} is monotone,
$\Omega_{\max}\le w$ if and only if every latent obeys $z_k\le t(w)$ with
$t(w)\triangleq\Phi^{-1}\!\bigl(F_{\Omega}(w)\bigr)$, so the \gls{cdf} of
$\Omega_{\max}$ is the $K$-variate Gaussian orthant probability
\begin{equation}
    F_{\Omega_{\max}}(w)=\Pr\{z_1\le t(w),\dots,z_K\le t(w)\}
    =\Phi_{\mat{R}}\!\bigl(t(w)\,\vec 1_K\bigr),
    \label{eq:omega_max_cdf}
\end{equation}
where $\Phi_{\mat{R}}(\cdot)$ is the zero-mean and unit-variance $K$-variate normal
\gls{cdf} with correlation $\mat{R}$ and $\vec 1_K$ the all-ones vector.
Equations~\eqref{eq:stat_law}
and~\eqref{eq:omega_max_cdf} constitute the statistical model from which
Section~\ref{sec:analysis} derives the threshold and the average \gls{bep}.

\section{Performance Analysis}
\label{sec:analysis}
The receiver designs its threshold from the conditional law~\eqref{eq:stat_law}
and averages the error probability over the selection
law~\eqref{eq:omega_max_cdf}. We state the optimal threshold, the average
\gls{bep}, and the diversity order over a single-antenna link.

For equiprobable bits, the minimum-error detector of~\eqref{eq:decision} is the
\gls{lrt}. By~\eqref{eq:stat_law}, the conditional density of
$\hat\sigma^2$ is Erlang with per-sample power
$\sigma_b^2=\sigma_w^2(1+\gamma_S b)$, so the log-likelihood ratio is
$\Lambda(\hat\sigma^2)=N\ln(\sigma_0^2/\sigma_1^2)
+N\hat\sigma^2\bigl(1/\sigma_0^2-1/\sigma_1^2\bigr)$, which is monotone in
$\hat\sigma^2$, so the test reduces to the single
threshold~\eqref{eq:decision}. With $\Lambda(\lambda_{\mathrm{th}})=0$,
$\sigma_0^2=\sigma_w^2$ and $\sigma_1^2=\sigma_w^2(1+\gamma_S)$, solving
for $\tau=\lambda_{\mathrm{th}}/\sigma_w^2$ gives
\begin{equation}
    \tau^\star(\gamma_S)
    = \frac{(1+\gamma_S)\,\ln(1+\gamma_S)}{\gamma_S},
    \label{eq:tau_opt}
\end{equation}
which is independent of the block length $N$. The receiver sets $\tau^\star$ per
block from $\gamma_S=\bar\gamma_S\,\Omega_{\max}$, in which $\Omega_{\max}=\Omega_{k^\star}$
is already available from the per-port power measurements used to select
$k^\star=\argmax_k|h_k|^2$, so no carrier recovery enters the threshold
and the operation stays non-coherent.
Evaluating the chi-square tails of~\eqref{eq:stat_law} at $\tau^\star$
yields the conditional \gls{bep},
\begin{equation}
    P_b(\gamma_S)
    = \tfrac{1}{2}\!\left[
        1 - F_{2N}\!\bigl(2N\tau^\star\bigr)
          + F_{2N}\!\left(\frac{2N\tau^\star}{1+\gamma_S}\right)
      \right],
    \label{eq:bep_cond}
\end{equation}
with $F_{2N}(\cdot)$ being the chi-square \gls{cdf} with $2N$ degrees of freedom, and the
average \gls{bep} follows over the selected power,
\begin{equation}
    \bar P_b
    = \int_0^\infty P_b\bigl(\bar\gamma_S\, w\bigr)\,
        \mathrm{d}F_{\Omega_{\max}}(w).
    \label{eq:bep_avg}
\end{equation}
The \gls{cdf} of $\Omega_{\max}$ in~\eqref{eq:bep_avg} is the Gaussian orthant
probability~\eqref{eq:omega_max_cdf}. Changing the integration variable
in~\eqref{eq:bep_avg} to the latent quantile $z$ through $w=F_{\Omega}^{-1}
(\Phi(z))$ reduces the average \gls{bep} to the one-dimensional quadrature
\begin{equation}
    \bar P_b = \int_{-\infty}^{\infty}
        P_b\!\bigl(\bar\gamma_S\,F_{\Omega}^{-1}(\Phi(z))\bigr)\,
        \mathrm{d}\Phi_{\mat{R}}\!\bigl(z\,\vec 1_K\bigr),
    \label{eq:bep_quad}
\end{equation}
in which the only factor that depends on the aperture is the orthant
$\Phi_{\mat{R}}(z\vec 1_K)$. \rev{Evaluating~\eqref{eq:bep_quad} as a
trapezoidal sum over $n_z$ latent nodes costs $O(n_zK^2M)$, independent of $N$
and of the \gls{snr}, with each node needing an inverse noncentral chi-square
quantile and a $K$-variate orthant from a quasi-Monte Carlo rule of $M$ lattice
points~\cite{genz1992}.}

\begin{proposition}
\label{prop:diversity}
At the Bayes threshold $\tau^\star$, the conditional \gls{bep}~\eqref{eq:bep_cond}
decays at high \gls{snr} as
\begin{equation}
    P_b(\gamma_S) \sim \frac{N^N}{2\,N!}\,(\ln\gamma_S)^N\,\gamma_S^{-N},
    \label{eq:cond_asym}
\end{equation}
so averaging over the selection law~\eqref{eq:omega_max_cdf} gives
$\bar P_b=O(\bar\gamma_S^{-d_K})$ with diversity order
\begin{equation}
    d_K = \min\!\left(N,\ \mu K_{\mathrm{eff}}\right),
    \label{eq:diversity}
\end{equation}
where $K_{\mathrm{eff}}=\vec 1_K^{\!\top}\mat{R}^{-1}\vec 1_K$ is the effective
number of independent ports set by the correlation~\eqref{eq:jakes}. It equals
$K$ for independent ports with $\mat{R}=\mat{I}_K$ the $K\times K$ identity matrix, collapses to $1$ for fully
correlated ports with $\mat{R}=\vec 1_K\vec 1_K^{\!\top}$, and can exceed $K$ when
the Jakes lobes make ports anti-correlated, as shown in Appendix~\ref{app:bound}.
The Bhattacharyya inequality, studied in~\cite{kailath1967divergence}, bounds the
conditional error as $P_b(\gamma_S)\le\tfrac{1}{2}\rho(\gamma_S)^N$, with
$\rho(\gamma_S)=2\sqrt{1+\gamma_S}/(2+\gamma_S)\to 2/\sqrt{\gamma_S}$ at high \gls{snr},
decays as $\rho(\gamma_S)^N\sim 2^N\gamma_S^{-N/2}$, an exponent $N/2$ that is half
the true detector order $N$ in~\eqref{eq:cond_asym}.
\end{proposition}

\ifappendices Proof: See Appendix~\ref{app:bound}. \qed \fi

The detector limits the diversity to $N$, whereas the fluid antenna contributes
$\mu K_{\mathrm{eff}}$, and the smaller of the two prevails. Setting $K=1$
recovers the single-antenna \gls{oodn} link with $d_1=\min(N,\mu)$, so selection raises
the diversity order by the factor $d_K/d_1=\min(K_{\mathrm{eff}},\,N/\mu)$, up to
the ceiling $N$. The logarithmic prefactor $(\ln\gamma_S)^N$ in~\eqref{eq:cond_asym}
makes the approach to this ceiling slow. Over the operational range
$\bar P_b\in[10^{-6},10^{-1}]$ the local slope stays close to the weaker
Bhattacharyya exponent $\min(N/2,\mu K_{\mathrm{eff}})$, and the true order
$\min(N,\mu K_{\mathrm{eff}})$ emerges only at vanishing \gls{bep}.

\section{Numerical Results}
\begin{table}[!t]
\centering
\caption{Simulation Parameters}
\label{tab:sim_params}
\footnotesize
\begin{tabular}{@{}ll@{}}
\hline
Parameter & Value \\
\hline
Modulation, detector & \glsentryshort{oodn}, energy detector \\
Block length $N$ & $8$ \\
Threshold & per-block Bayes $\tau^\star(\gamma_S)$,~\eqref{eq:tau_opt} \\
Fading, \glsentryshort{los} $(\kappa,\mu)$ & $(3.53,\,1.32)$ \\
Fading, \glsentryshort{nlos} $(\kappa,\mu)$ & $(1.08,\,0.84)$ \\
Port count $K$ & $\{1,2,4,8\}$ \\
Aperture $W$ & $\{0.15,0.3,0.6,1.5\}\lambda$ \\
Spatial correlation & Jakes $\mat{R}$~\eqref{eq:jakes}, Gaussian copula \\
\glsentryshort{snr} range & $0$ to $24$~dB \\
\glsentryshort{mc} per point & adaptive to $300$ errors, up to $10^9$ bits \\
\hline
\end{tabular}
\end{table}

We validate the analysis by \gls{mc} simulation of the \gls{oodn} link under the parameters of Table~\ref{tab:sim_params}, with two $\kappa$-$\mu$ operating points, \gls{los} and \gls{nlos}, fitted to measured millimeter-wave data~\cite{reis_2019}. With $N=8$, the single-port
reference is fading-limited at both operating points because $\mu<N/2$, whereas selection
raises the fading exponent to $\mu K_{\mathrm{eff}}$ and turns the link detector-limited
from $K=4$ under \gls{los} and $K=8$ under \gls{nlos}. Each \gls{mc} point runs adaptively to $300$ bit
errors. Correlated ports are drawn from the Gaussian copula~\eqref{eq:copula} on the full Jakes
matrix~\eqref{eq:jakes}, exactly the channel the theory curves integrate through
the \gls{cdf}~\eqref{eq:omega_max_cdf} of $\Omega_{\max}$. \rev{The receiver
activates the single port of largest gain, $k^\star=\argmax_k|h_k|^2$, in every
figure, so the law of the selection statistic $\Omega_{\max}$ is set by the
correlation matrix $\mat{R}$ alone.} In all figures,
lines are theoretical and markers \gls{mc}, with solid lines and filled markers for
\gls{los} and dashed lines and open markers for \gls{nlos}.

\begin{figure}[!t]
    \centering
    \includegraphics{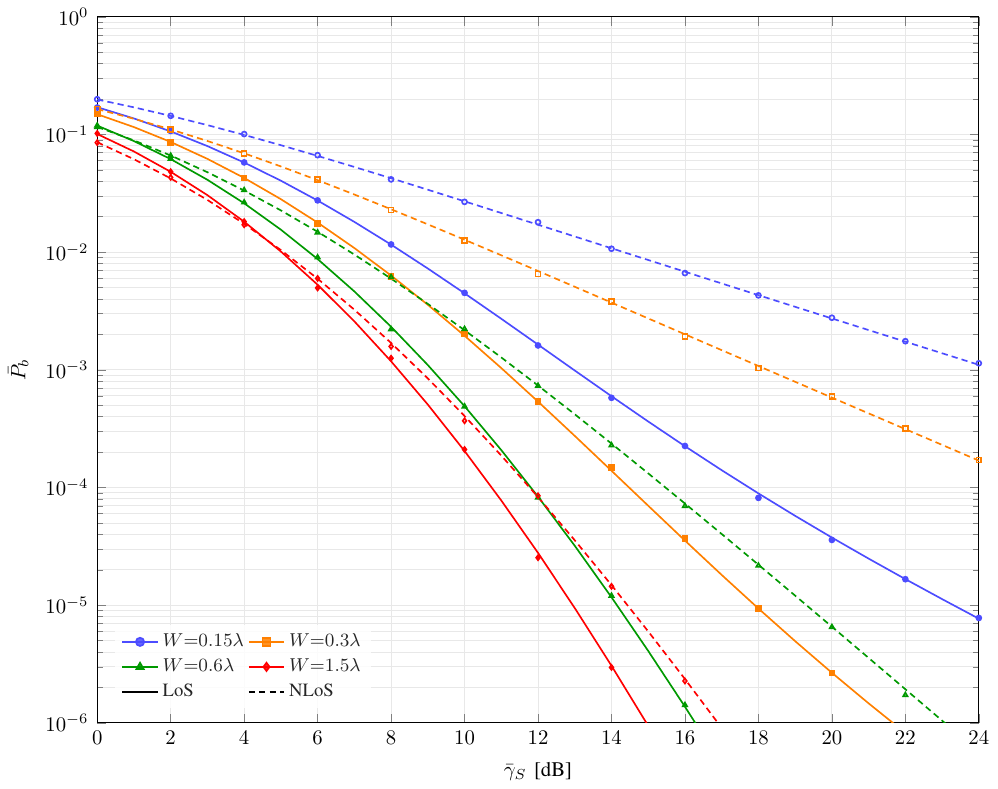}
    \caption{Average \gls{bep} versus transmit \gls{snr} for $K=4$ at several apertures $W$,
    with $N=8$, for the \gls{los} and \gls{nlos} cases.}
    \label{fig:bep_snr_W}
\end{figure}

Fig.~\ref{fig:bep_snr_W} fixes $K=4$ and overlays the \gls{bep}-versus-\gls{snr}
waterfall for several apertures $W$. At the smallest aperture the ports are
almost fully correlated, and as $W$ grows they decorrelate and the waterfall
steepens, spanning the full range of diversity orders available at this $K$.

\begin{figure}[!t]
    \centering
    \includegraphics{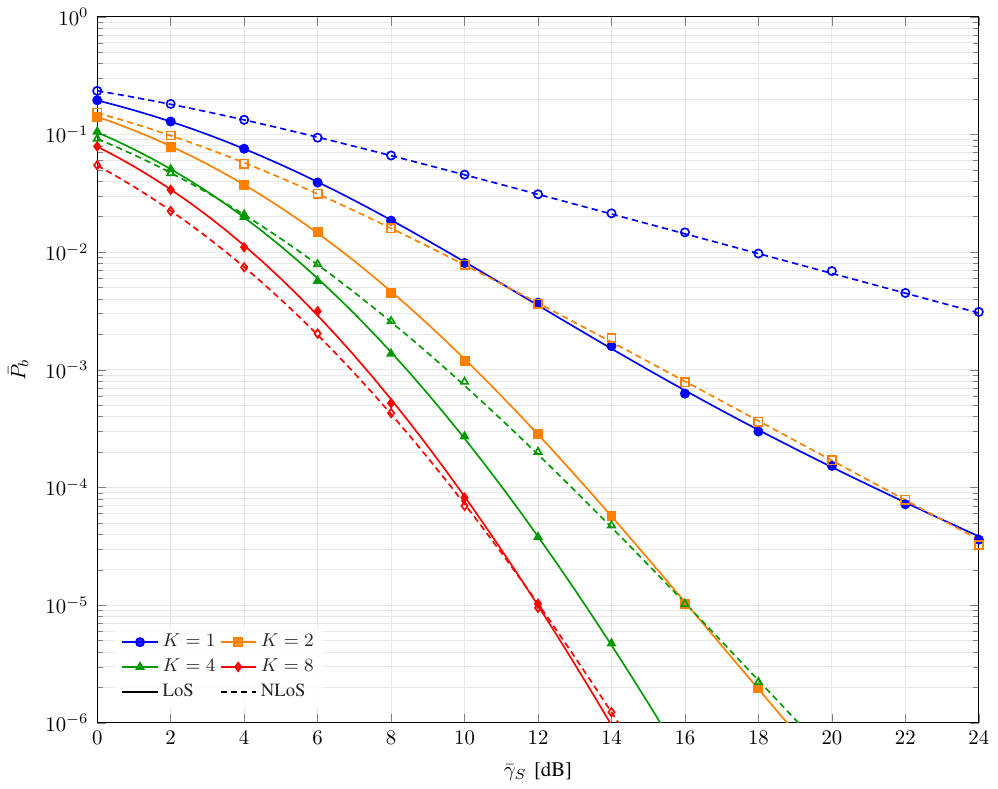}
    \caption{Average \gls{bep} versus transmit \gls{snr} for the \gls{los} and \gls{nlos}
    cases under independent port selection $\mat{R}=\mat{I}_K$ at $N=8$.}
    \label{fig:bep_snr}
\end{figure}

Fig.~\ref{fig:bep_snr} shows the average \gls{bep} versus the transmit \gls{snr}
$\bar\gamma_S$ for $K\in\{1,2,4,8\}$ under the independent-selection idealization
$\mat{R}=\mat{I}_K$. The theoretical curves from~\eqref{eq:bep_avg} match the
\gls{mc} markers over the whole range. Setting $K=1$ recovers the single-antenna
\gls{oodn} link, and increasing $K$ steepens the waterfall as the extra independent
ports add the spatial diversity of Proposition~\ref{prop:diversity}.

\begin{figure}[!t]
    \centering
    \includegraphics{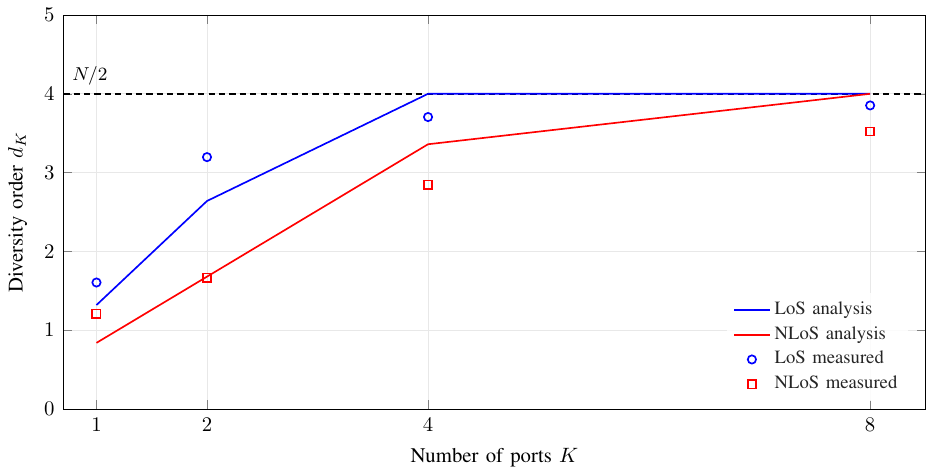}
    \caption{Operational diversity order versus number of ports $K$ under
    independent selection $\mat{R}=\mat{I}_K$, with $N=8$, for the \gls{los} and
    \gls{nlos} cases.}
    \label{fig:diversity}
\end{figure}

Finally, Fig.~\ref{fig:diversity} verifies the operational slope of
Proposition~\ref{prop:diversity}. For each $K$ the operational Bhattacharyya exponent
$\min(N/2,\mu K_{\mathrm{eff}})$ is compared with the local
slope of the exact \gls{bep}~\eqref{eq:bep_avg} at $\bar P_b=10^{-5}$. Both cases track this exponent and approach the ceiling
$N/2$. The asymptotic order $\min(N,\mu K_{\mathrm{eff}})$ of~\eqref{eq:diversity} is
deferred to much lower \gls{bep} by the logarithmic prefactor of~\eqref{eq:cond_asym}.
\rev{The correlated case $\mat{R}\neq\mat{I}_K$ is left for future work.}

\section{Conclusion}
This letter \rev{integrated} \gls{oodn} modulation with a \gls{fas} and developed a
unified framework for its non-coherent detection over \gls{awgn} and
$\kappa$-$\mu$ fading. Retaining the full Jakes correlation through a Gaussian
copula, we derived the \rev{Bayes threshold, the average \gls{bep}, and a
diversity order} that is $\min(N,\mu K_{\mathrm{eff}})$ asymptotically and follows the
operational Bhattacharyya exponent $\min(N/2,\mu K_{\mathrm{eff}})$ over the
practical \gls{bep} range validated by Monte Carlo simulation. \rev{Port selection
substantially improves the reliability of \gls{oodn}} while preserving its
low-complexity, phase-noncoherent operation, making it attractive for
energy-constrained \gls{iot} and machine-type devices.

\ifappendices
\appendices
\section{Proof of Proposition~\ref{prop:diversity}}
\label{app:bound}
At the Bayes threshold $\tau^\star=(1+\gamma_S)\ln(1+\gamma_S)/\gamma_S$
of~\eqref{eq:tau_opt}, the conditional \gls{bep}~\eqref{eq:bep_cond} splits into a
false-alarm and a miss event of the scaled statistic~\eqref{eq:stat_law}, as
\begin{equation}
    P_b(\gamma_S)
    = \tfrac{1}{2}\bigl[\,1-F_{2N}(2N\tau^\star)\bigr]
    + \tfrac{1}{2}\,F_{2N}\!\left(\frac{2N\tau^\star}{1+\gamma_S}\right),
    \label{eq:pb_decomp}
\end{equation}
with $F_{2N}(\cdot)$ the chi-square \gls{cdf} of order $2N$. Evaluating the two terms
as $\gamma_S\to\infty$, the miss event sets the decay.

Writing $L=\ln(1+\gamma_S)$, the miss argument $2N\tau^\star/(1+\gamma_S)=2NL/\gamma_S$
tends to zero, so the lower incomplete gamma expansion $\gamma(N,x)\sim x^N/N$ as
$x\to0$ gives
\begin{align}
    F_{2N}\!\left(\frac{2N\tau^\star}{1+\gamma_S}\right)
    &= \frac{\gamma\!\bigl(N,\,NL/\gamma_S\bigr)}{\Gamma(N)} \notag\\
    &\sim \frac{1}{\Gamma(N)}\,\frac{1}{N}\left(\frac{NL}{\gamma_S}\right)^{\!N} \notag\\
    &= \frac{1}{N!}\left(\frac{N\ln\gamma_S}{\gamma_S}\right)^{\!N}.
    \label{eq:miss_asym}
\end{align}
The false-alarm threshold instead grows, $2N\tau^\star\sim 2NL\to\infty$, so the
upper chi-square tail gives
\rev{$1-F_{2N}(2N\tau^\star)=\Gamma(N,N\tau^\star)/\Gamma(N)
=O\bigl((\ln\gamma_S)^{N-1}\gamma_S^{-N}\bigr)$,}
which carries one fewer power of $\ln\gamma_S$ than~\eqref{eq:miss_asym} and is
therefore subdominant. Substituting both terms into~\eqref{eq:pb_decomp}
\rev{reproduces the conditional decay~\eqref{eq:cond_asym}}, so the detector
contributes exponent $N$.

It remains to average this exponent over the law of the selected gain. As $w\to0$ the latent
quantile $t(w)=\Phi^{-1}(F_{\Omega}(w))\to-\infty$, and the Gaussian orthant
probability~\eqref{eq:omega_max_cdf} obeys the large-deviations tail
$\Phi_{\mat{R}}(t\vec 1_K)\sim[\Phi(t)]^{\,\vec 1_K^{\!\top}\mat{R}^{-1}\vec 1_K}$,
obtained by minimizing the quadratic form $\tfrac12\vec x^{\!\top}\mat{R}^{-1}\vec x$
over the orthant $\vec x\le t\vec 1_K$: when $\mat{R}^{-1}\vec 1_K\ge\vec 0$ the
Karush--Kuhn--Tucker conditions place the minimizer at $\vec x=t\vec 1_K$, which
holds for the correlation matrices considered here.
With $F_{\Omega}(w)=\Phi(t(w))\sim w^{\mu}$, the \gls{cdf} of $\Omega_{\max}$ therefore
behaves as $F_{\Omega_{\max}}(w)\sim w^{\mu K_{\mathrm{eff}}}$ and
$f_{\Omega_{\max}}(w)\sim\mu K_{\mathrm{eff}}\,w^{\mu K_{\mathrm{eff}}-1}$, with
$K_{\mathrm{eff}}=\vec 1_K^{\!\top}\mat{R}^{-1}\vec 1_K$. Inserting the
conditional decay $P_b(\bar\gamma_S w)=O\bigl((\bar\gamma_S w)^{-N}\bigr)$
into~\eqref{eq:bep_avg}, the integral is set by the dominant balance between the
small-$w$ mass of the $\Omega_{\max}$ density $w^{\mu K_{\mathrm{eff}}-1}$ and the
$w^{-N}$ growth of the conditional error, and a Laplace argument gives
\begin{equation}
    \bar P_b
    = \int_0^\infty P_b(\bar\gamma_S w)\,\mathrm{d}F_{\Omega_{\max}}(w)
    = O\!\bigl(\bar\gamma_S^{-\min(N,\,\mu K_{\mathrm{eff}})}\bigr),
    \label{eq:avg_asym}
\end{equation}
the exponent being the smaller of the detector exponent $N$ and the fading
exponent $\mu K_{\mathrm{eff}}$, which is the diversity order~\eqref{eq:diversity}.

The Bhattacharyya bound follows the same argument but is weaker by a factor of two
in the exponent. Under hypothesis $b$, the samples $|x_{k^\star,n}|^2$ are
i.i.d.\ exponential of mean $\sigma_b^2$, with $\sigma_0^2=\sigma_w^2$ and
$\sigma_1^2=\sigma_w^2(1+\gamma_S)$. Writing $f_b(x)=\sigma_b^{-2}e^{-x/\sigma_b^2}$
for the per-sample density under hypothesis $b$, the Bhattacharyya coefficient
integrates to
\begin{align}
    \int_0^\infty\!\sqrt{f_0(x)f_1(x)}\,\mathrm{d}x
    &= \frac{1}{\sigma_0\sigma_1}\int_0^\infty\!
        \exp\!\left[-\frac{x}{2}\!\left(\frac{1}{\sigma_0^2}
            +\frac{1}{\sigma_1^2}\right)\right]\mathrm{d}x \notag\\
    &= \frac{2\sigma_0\sigma_1}{\sigma_0^2+\sigma_1^2}
     = \frac{2\sqrt{1+\gamma_S}}{2+\gamma_S}
     = \rho(\gamma_S).
    \label{eq:bhat_coef}
\end{align}
Independence across the $N$ samples raises this to $\rho(\gamma_S)^N$, so
$P_b(\gamma_S)\le\tfrac{1}{2}\rho(\gamma_S)^N$, and at high \gls{snr}
\begin{align}
    \rho(\gamma_S)^N
    = \left[\frac{4(1+\gamma_S)}{(2+\gamma_S)^2}\right]^{N/2}
    \longrightarrow \left(\frac{4}{\gamma_S}\right)^{\!N/2},
    \label{eq:rho_asym}
\end{align}
an exponent of only $N/2$, half the true value. Averaging over $\Omega_{\max}$ as
above replaces $N$ by $\mu K_{\mathrm{eff}}$, so the operational exponent is
$\min(N/2,\mu K_{\mathrm{eff}})$. \qed

\iflongversion
\section{Proof of Proposition~\ref{prop:arraygain}}
\label{app:mixture}
The generalized Marcum-$Q$ admits the series
$Q_\mu(a,b)=\sum_{j\ge0}\frac{e^{-a^2/2}(a^2/2)^j}{j!}
\Gamma(\mu+j,\,b^2/2)/\Gamma(\mu+j)$. With $a^2/2=\kappa\mu$ and
$b^2/2=\mu(1+\kappa)w$, the \gls{cdf}~\eqref{eq:kappamu_cdf} becomes the Poisson mixture
of Gamma laws
\begin{equation}
    F_{\Omega}(w) = \sum_{j=0}^{\infty} p_j\,
        \frac{\gamma\!\bigl(\mu+j,\ \mu(1+\kappa)w\bigr)}{\Gamma(\mu+j)},
    \qquad p_j=\frac{e^{-\kappa\mu}(\kappa\mu)^{j}}{j!},
    \label{eq:mixture}
\end{equation}
with $\gamma(\cdot,\cdot)$ the lower incomplete gamma function. Differentiating and
integrating $w^{-N/2}$ against the $j$-th Gamma density gives
\begin{align}
    \E\bigl[\Omega^{-N/2}\bigr]
    &= \sum_{j=0}^{\infty} p_j\,
       \frac{[\mu(1+\kappa)]^{\mu+j}}{\Gamma(\mu+j)}
       \int_0^\infty\! w^{\mu+j-1-\frac{N}{2}}
       e^{-\mu(1+\kappa)w}\,\mathrm{d}w \notag\\
    &= [\mu(1+\kappa)]^{N/2}\sum_{j=0}^{\infty} p_j\,
       \frac{\Gamma\!\bigl(\mu+j-\tfrac{N}{2}\bigr)}{\Gamma(\mu+j)} \notag\\
    &= [\mu(1+\kappa)]^{N/2}
       \frac{\Gamma\!\bigl(\mu-\tfrac{N}{2}\bigr)}{\Gamma(\mu)}
       \sum_{j=0}^{\infty}
       \frac{\bigl(\mu-\tfrac{N}{2}\bigr)_j}{(\mu)_j}
       \frac{(\kappa\mu)^{j}}{j!}\,e^{-\kappa\mu} \notag\\
    &= [\mu(1+\kappa)]^{N/2} e^{-\kappa\mu}
       \frac{\Gamma\!\bigl(\mu-\tfrac{N}{2}\bigr)}{\Gamma(\mu)}\,
       {}_1F_1\!\bigl(\mu-\tfrac{N}{2};\mu;\kappa\mu\bigr),
    \label{eq:negmom_der}
\end{align}
where the third line uses
$\Gamma(\mu+j-\tfrac{N}{2})/\Gamma(\mu+j)=
[\Gamma(\mu-\tfrac{N}{2})/\Gamma(\mu)]\,(\mu-\tfrac{N}{2})_j/(\mu)_j$ with
$(\cdot)_j$ the Pochhammer symbol and the last line summing the Kummer series,
both standard special-function identities~\cite{gradshteyn2014}. This
proves~\eqref{eq:negmom}. Substituting $\rho(\gamma_S)^N\to(4/\gamma_S)^{N/2}$
into the conditional bound $P_b\le\tfrac{1}{2}\rho(\gamma_S)^N$ and averaging with
$K=1$ then gives the asymptote~\eqref{eq:bep_asym_nofas}. \qed
\fi
\fi

\printbibliography

\end{document}